# The Mauna Kea Observatories Near-Infrared Filter Set. I: Defining Optimal 1–5 µm Bandpasses

D. A. Simons

*Gemini Observatory, Northern Operations Center, 670 A`ohoku Place, Hilo HI 96720;*

`dsimons@gemini.edu`

A. T. Tokunaga

*University of Hawaii, Institute for Astronomy, 2680 Woodlawn Drive, Honolulu, HI 96822;*

`tokunaga@ifa.hawaii.edu`

Editorial correspondence to:   Douglas Simons
Gemini Observatory
670 A`ohoku Place
Hilo, HI.   96720

**ABSTRACT**

A new MKO-NIR infrared filter set is described, including techniques and considerations given to designing a new set of bandpasses that are useful at both mid- and high-altitude sites. These filters offer improved photometric linearity and in many cases reduced background, as well as preserve good throughput within the *JHKLM* atmospheric windows. MKO-NIR filters have already been deployed within a number of instruments around the world as part of a filter consortium purchase to reduce the unit cost of filters. Through this effort we hope to establish, for the first time, a *single standard* set of infrared filters at as many observatories as possible.

1. INTRODUCTION

Rigorous standardization of infrared filters used at various sites has never been achieved. Since the establishment of the JKLM system by Johnson (1966) there have been numerous infrared photometric systems and lists of standard stars published, including the Anglo-Australian Observatory (AAO; Allen & Cragg 1983), Arcetri Observatory (Hunt et al. 1998), California Institute of Technology (CIT; Elias et al. 1982), European Southern Observatory (ESO; van der Bliek et al. 1996), Las Companas Observatory (LCO; Persson et al. 1998), United Kingdom Infrared Telescope (UKIRT; Hawarden et al. 2001), Mount Stromlo and Siding Spring Observatory (MSSSO; McGregor 1994), and South Africa Astronomical Observatory (SAAO; Carter 1990). See also the list of other photometric systems in Glass (1999) and Tokunaga (2000). Considerable effort has been made to calibrate various filter systems used at different observatories. For example Elias et al. (1982) and Bessel & Brett (1988) produced calibration relations between various photometric standard star systems, achieving typically ~0.02 mag uncertainties in color transformations.

None of these observatories use a common infrared filter set since there has not been a single continual source for astronomical infrared filters, yet alone an agreement among observatories as to the optimum filter set. As infrared observational techniques have become more sophisticated and accurate, attention to systematic photometric errors is becoming more important. Manduca and Bell (1979), Milone (1989) and Young, Milone, & Stagg (1994) point out the many problems with obtaining precision photometry with the relatively wide infrared filters in use to date.

Water vapor plays an important role in determining atmospheric extinction, since it can vary significantly from night-to-night or between observing sites. Filters that are used commonly across many sites must be designed to be relatively immune to water vapor contamination, thereby driving filter bandwidths to be fairly narrow. This was discussed in detail by Young et al. (1994).

The Mauna Kea Observatories Near-Infrared (MKO-NIR) filters described in the following pages represent a compromise between the competing factors of throughput and photometric performance. They are perhaps the first widely used filters that have been designed for use at both high-altitude sites like Mauna Kea and mid-altitude sites like Palomar or Kitt Peak. While not perfectly optimized for either mid- or high-altitude sites, they offer good throughput and linearity under a variety of conditions, and in general minimize sky background flux without significant sacrifice in terms of overall throughput. Thus they are reasonably well designed for optimum broadband signal-to-noise ratios. Many of the new MKO-NIR bandpasses are fairly close to those already commonly used today, but some, particularly the *J* bandpass, are considerably narrower due to the historically very poor design of this filter, which includes portions of the atmosphere with little to no transparency and large amounts of OH emission.

The MKO-NIR filters are further described in "The Mauna Kea Observatories Near–Infrared Filter Set II: Specifications for a New *JHKL´M´* Filter Set for Infrared Astronomy" (Tokunaga, Simons, & Vacca 2001), which discusses in detail the fabrication specifications for these filters. They also discuss the performance of this filter set when extrapolated to zero air mass and demonstrate their superior performance under varying water vapor conditions compared to other filters commonly used today. After Simons led the effort to define new bandpasses as for the Gemini Telescope Project, Tokunaga led the effort to organize a large consortium purchase of these filters. Our motivation in defining a new filter set arose from our need to provide filters for instrumentation at the Gemini and Subaru Telescopes on Mauna Kea. Consistent with the top-level design philosophies of these new 8 m telescopes, which exploit natural site conditions as much as possible, it was decided to reexamine the central wavelengths and bandpasses of filters to minimize the effects of the atmosphere. It was also hoped that through a consortium purchase of these filters, which significantly reduced the cost per user, wide acceptance of these filters would lead to standardization throughout the infrared community.

## 2. ANALYSIS

Theoretical filter performances were estimated through the use of a variety of model atmospheric absorption and emission data. Discussed in this section are various performance metrics that were used to judge filter performance against a number of factors.

### 2.1. Input Model Data

Filter bandpasses were evaluated by using MathCAD and a set of model atmospheres provided by a number of sources. For Mauna Kea, a 1–6 µm model atmosphere generated by MODTRAN (Abreau & Anderson 1996) was kindly provided by G. Milone. The model consists of atmospheric absorption at air masses of 1.0, 1.5, 2.0, 2.5, and 3.0 for an assumed 1 mm precipitable water vapor content at a 4200 m altitude location in the mid-tropics. Accordingly, it represents typical atmospheric conditions over Hawaii. This absorption model was merged with a night-sky emission model based upon the work of P. Roche and A. Glasse (1990; private communication) to generate integrated sky flux levels for the various bandpasses under consideration. Finally, S. Lord at the NASA Jet Propulsion Laboratory kindly provided a model atmosphere for a 2 km mid-altitude site, also at 1.0, 1.5, 2.0, 2.5, and 3.0 air masses. The MODTRAN, JPL, and sky emission models all had spectral resolutions of $\lambda/\Delta\lambda \sim 10^4$. No emission model was available for the mid-altitude site for this analysis, hence all sky backgrounds are based upon relatively cold Mauna Kea sky conditions.

### 2.2. Model Data Processing

Theoretical filter curves were generated by convolving a simple Gaussian function with a boxcar function of specified width. The convolution of these functions, with degrees of freedom specified in terms of bandpass width, central wavelength, peak transmission, and roll-off, offered good control over pertinent model factors and simulated the transmission curves of real filters fairly well. All filters were assumed to have a peak transmission of 85%, which is consistent with what is typically achievable at commercial coating facilities.

Model bandpasses were then generated by first estimating a central wavelength for each atmospheric window to define the central wavelength, $\lambda_c$. Given the very strong absorption that characterizes the edges of most atmospheric windows, picking a mean wavelength to center the filter bandpass is a reasonable means of accelerating processing time, compared to letting $\lambda_c$ run as free parameter. Likewise, a fixed roll-off was used for each series of bandpasses. It was derived by empirically comparing model bandpasses with what is typically achieved in the production of commercial near-infrared filters. Again, the primary intent of using a fixed instead of variable roll-off was to speed up processing time. Then, for a fixed central wavelength and roll-off, the filter bandpasses were allowed to expand in increments, $\Delta\lambda$, of typically ~0.03 µm for *J*, *H*, and *K,* and ~0.08 µm for *L´* and *M´*. Table 1 lists the $\lambda_c$ adopted for each atmospheric window. For each $\lambda_c \pm \Delta\lambda$ increment the following performance parameters were calculated:

1. The zenith absorption, $\tau_0$, for both the Mauna Kea and a mid-altitude 2 km site, defined as

$$\tau_0 = \frac{1}{\Delta\lambda} \int_{\lambda_1}^{\lambda_2} f(\lambda)\tau(\lambda)d\lambda \qquad (1)$$

where *f(λ)* is the filter function, *τ(λ)* is the atmospheric absorption at zenith, $\lambda_1$ and $\lambda_2$ are the ~10% transmission points in the filter profile and $\Delta\lambda = \lambda_2 - \lambda_1$. The parameter $\tau_0$ is therefore the mean value of the product *f(λ)τ(λ)* across $\Delta\lambda$ and gauges when further broadening of a proposed bandpass no longer leads to increased flux through a filter.

2. Using *τ(λ)* data for the other air masses, the mean absorptions $\tau_1$, $\tau_2$, $\tau_3$, and $\tau_4$ corresponding to 1.5, 2.0, 2.5, and 3.0 air masses were also calculated, using the same approach defined in equation (1). From there, a $\chi^2$ metric for the nonlinearity of photometric extinction, as a function of air mass $\tau_i$, was determined. This provided an estimate of the photometric error in magnitudes associated with each trial bandpass.

3. The theoretical sky plus telescope background in terms of magnitudes per arcsec$^2$ and photons arcsec$^{-2}$ sec$^{-1}$ at the Gemini focal plane was also calculated for each model bandpass.

The calculation assumed a telescope plus instrument emissivity of 3%, with the telescope pointing at zenith and having an ambient temperature of 0° C.

### 2.3. Model Output

Figure 1 shows a typical prototype filter bandpass used in the investigation of the *K* band. Also shown is the Mauna Kea zenith atmospheric absorption from ~1.9 to 2.6 µm. Figure 2 shows the corresponding extinction linearity analysis for this bandpass. The line represents a least-squares fit to the extinction measured between 1 and 3 air masses. Figure 3 shows the Mauna Kea sky emission for this spectral region, which was used in conjunction with the filter bandpass to estimate sky background levels for each $\Delta\lambda$ increment in the analysis. In the end a large model database was generated for each of the atmospheric windows from 1 to 5 µm. They are plotted and discussed in section 3.

### 3. THE MKO-NIR BANDPASSES

A number of optimization strategies can be used to define near-infrared broadband filters. For example, Young et al. (1994) conducted a detailed study of the *JHKLMNQ* bands by defining a figure of merit for the linearity of extinction curves for various sites under various water vapor conditions. Bandpasses were optimized primarily on the basis of reproducibility and transformability of photometric measurements across various sites.

The approach adopted here was to assess extinction linearity, throughput, and background flux for possible bandpasses in a relatively simple manner. Unlike Young et al.

(1994), total throughput (atmosphere + filter) was heavily weighted in the optimization process, not just photometric performance, on the assumption that most astronomical applications demand peak signal-to-noise even if it means a few percent degradation in photometric accuracy. The result is a set of bandpasses that are considerably broader than the triangle-shaped highly optimized bandpasses derived by Young et al. (1994), and that have increased throughput but still fairly good photometric performance (see Fig. 4).

Table 2 lists the filter sets that have been used for many years at several observatories on Mauna Kea, including those in NSFCAM (IRTF), IRCAM3 (UKIRT), and Redeye (CFHT). It also lists the Barr Associates standard astronomy set to illustrate filter commonality between these observatories. The MKO-NIR filters were also compared with stock *JHKLL´M* filters commonly manufactured by Barr Associates to quantify the benefit the new MKO-NIR filters would provide compared with this commonly used set.

Plotted in Figures 5 through 11 are standard Barr and the new MKO-NIR filter bandpasses for the *J, H, K, L´,* and *M´* windows. All filter curves have been plotted with the Mauna Kea model atmosphere for 1.0 air masses to illustrate how the filters "fit" within windows. Also shown at the bottom of each figure is an estimate of the photometric nonlinearity $\sigma$ (in millimag) and $\tau_0$ as a function of filter bandpass for Mauna Kea and a 2 km altitude site. A vertical dashed line corresponds to the MKO-NIR bandpass at the ~10% transmission level in the filter profile. In general the final bandpass was selected on the basis of when total throughput ($\tau_0$) begins to fall and photometric error ($\sigma$) over the 1.0–3.0 air mass range begins to rise rapidly with increasing $\Delta\lambda$. Also shown for each window is the Mauna Kea zenith atmosphere plus telescope emission. This emission plot is not included in the *J*-band plots because the model only covers $\lambda > 1.45$ µm. Here are specific comments about the filters for each wavelength band:

*J* band: The commonly used *J*-band filter is clearly a poor match to the atmosphere. Nearly 50% of the bandpass covered by the Barr filter is contaminated by water vapor lines, leading to a ~5× greater photometric error contribution than the MKO-NIR bandpass. Furthermore, the *J* window is heavily contaminated with OH emission lines that are needlessly passed through the oversized Barr filter bandpass. The net result is significantly increased sky background (~0.5 mag) and photometric error over the MKO-NIR bandpass.

*H* band: The Barr *H* filter intrudes slightly into the water vapor band near 1.80 µm but is otherwise a reasonably good fit to this window. The photometric error rises dramatically for bandpasses greater than ~0.30 µm, leading to the MKO-NIR *H* filter that has a width slightly under 0.30 µm.

*K* band: The commonly used K´ filter (Wainscoat & Cowie 1992) includes shorter wavelengths than $K_s$, hence it is more susceptible to water vapor impacting performance than $K_s$. A different filter dubbed $K_{long}$ was considered as part of the analysis. It takes advantage of Gemini's low emissivity and avoids the deep $CO_2$ absorption lines at 2.01 and 2.06 µm. No significant increase in sky background would result from such a bandpass while throughput and extinction linearity would be improved over $K_s$ and K´. The overall performance of $K_{long}$ is similar to the MKO-NIR *K* filter, since both avoid the deep $CO_2$ lines and have comparable width. Since this filter is tuned for low emissivity telescopes, the advantages it theoretically offers may not be easily applicable to telescopes at warmer sites. As a result it was not adopted as an element of the new MKO-NIR standard filter set.

$L´$ band: Since the red (high-emissivity) end of the ~4 µm window is also clear of telluric lines, simultaneously optimizing this bandpass for low background and good photometric performance is difficult. The MKO-NIR $L´$ bandpass corresponds to the transmission turnover point for both the Mauna Kea and 2 km sites hence it is optimized for throughput at both sites. The 2 km σ increases rapidly beyond $\Delta\lambda$ ~ 0.5 µm, while the MKO σ increases much more slowly. In this case, the MKO-NIR bandpass was chosen to be $\Delta\lambda = 0.7$ µm to favor high throughput at both sites while giving good photometric linearity on Mauna Kea.

$M´$ band: The IRTF $M´$ filter suffers from several absorption lines, leading to an estimated photometric error of ~8 millimag when used on Mauna Kea and ~10 millimag on a 2 km altitude site. The MKO-NIR filter is shifted significantly bluer, into a more transparent region within the $M$ window. This region also offers significantly lower background, thereby boosting the overall signal-to-noise ratio. The MKO-NIR filter is actually close to the nb$M$ filter in use at UKIRT and has a theoretical photometric error contribution of ~7 millimag on Mauna Kea. Like the $L´$ filter, a significant difference exists in the photometric performance of such filters when used on Mauna Kea and a 2 km altitude site, making it difficult to select a single bandpass that is well suited for both mid- and high-altitude sites.

# 4. RESULTS

Table 3 summarizes key performance levels and bandpass definitions for the commonly used Barr and new MKO-NIR filters. All but one MKO-NIR bandpass (L´) provides improved photometry and reduced background over the Barr set with no loss in effective throughput. In some cases, the difference between existing filters and new ones is small but in all cases the MKO-NIR bandpasses meet or exceed the performance of commonly used near-infrared filters. In the case of the *J* filter, the MKO-NIR bandpass will no doubt lead to significantly improved photometry and a better signal-to-noise ratio (due to the ~0.5 mag reduced background), with no real loss in effective throughput since so much of the Barr *J* filter is heavily obscured by water vapor absorption. In general, not surprisingly, lower photometric errors will be achieved on Mauna Kea than lower altitude sites, but at least at *J*, *H*, and *K*, the differences will be reduced for the MKO-NIR bandpasses. The *L´* filter in particular represents a difficult set of trades. The ~4 µm window is relatively clear of telluric lines at the red (high emissivity) end of the window, so simultaneously minimizing the effect of atmospheric absorption and emission is difficult. Ultimately the signal-to-noise ratio achieved for science targets is a function of the color of the target, which can range from stellar (peak flux short of ~4 µm) to dust (peak flux beyond ~4 µm). Biasing the location of $\lambda_c$ toward the blue end of this window to cut the background will therefore not necessarily lead to the highest signal-to-noise ratio for all sources. The best compromise between competing factors is to use a bandpass wide enough to give good throughput at both mid- and high-altitude sites and good photometric linearity for at least Mauna Kea. When compared with the Barr *L´* filter at 2 km, the MKO-NIR filter has only ~0.6 millimag greater theoretical error, which in typical 3–5 µm observations is negligible. No suggested changes are offered here to the *L* and *M* Barr filters, because these obsolete bandpasses are logically replaced by the *L*´ and *M*´ filters anyway. This analysis favors an *M*´ filter closely resembling that already in use at UKIRT (nb*M* in IRCAM3), which offers a relatively low background and good extinction linearity. A new specialty *K* filter is proposed, dubbed $K_{long}$. It offers on Gemini improved throughput and photometry over $K_s$ or *K*´ with no increase in background, no doubt due to the low emissivity of the Gemini telescopes. However, the gains of such a filter may not be achievable on other telescopes with higher emissivities.

## 5. SUMMARY

The new MKO-NIR filter set offers many advantages over past and current filters commonly in use at observatories. While they are to first order similar to most broadband filters in use today, they have been designed to work well at both high- and mid-altitude sites and to offer good throughput and photometric stability. Among all the MKO-NIR filters, the *J* band filter represents the most significant change from current popular filters. The MKO-NIR *J* band filter bandpass has been tuned to closely track the ~1.2 µm atmospheric window, thereby simultaneously improving photometric linearity *and* preserving total throughput while significantly cutting OH emission across this bandpass.

# 6. ACKNOWLEDGEMENTS

The authors wish to thank Pat Roche, Alistair Glasse, Gene Milone, and Steve Lord for providing the model data required to perform the analysis outlined in this manuscript. The analysis used to generate the new MKO-NIR filter set would not have been possible without their generous contributions of model data. We also thank Sandy Leggett and Tim Hawarden for valuable advice throughout the process of defining this filter set.

This research was supported by the Gemini Observatory, which is operated by the Association of Universities for Research in Astronomy, Inc., under a cooperative agreement with the NSF on behalf of the Gemini partnership: the National Science Foundation (United States), the Particle Physics and Astronomy Research Council (United Kingdom), the National Research Council (Canada), CONICYT (Chile), the Australian Research Council (Australia), CNPq (Brazil) and CONICET (Argentina).

7. REFERENCES

Abreau, L. W. & Anderson, G. P. 1996, "The MODTRAN 2/3 Report and LOWTRAN 7 Model", Prepared by ONTAR Corporation for PL/GPOS

Allen, D.A., & Cragg, T.A. 1983, MNRAS, 203, 777

Bessell, M. S. & Brett, J. M. 1988, PASP, 100, 1134

Carter, B. S. 1990, MNRAS, 242, 1

Elias, J. H., Frogel, J. A., Matthews, K., & Neugebauer, G. 1982, AJ, 87, 1029; erratum 1982, AJ, 87, 1893

Glass, I. S. 1999, Handbook of Infrared Astronomy, Cambridge: Cambridge Univ. Press

Hawarden, T.G., Leggett, S.K., Letawsky, M.B., Ballantyne, D.R., & Casali, M.M. 2001, MNRAS, in press.

Hunt, L. K., Mannucci, F., Testi, L., Migliorini, S., Stanga, R. M., Baffa, C., Lisi, F., & Vanzi, L. 1998, AJ, 115, 2594; erratum 1999, AJ, 119, 985

Manduca, A. & Bell, R.A. 1979, PASP, 91, 848

McGregor, P.J. 1994, PASP, 106, 508

Milone, E. F. 1989, Proceeding of IAU Commissions 25 & 9, Springer-Verbig:NY, 79

Persson, S.E., Murphy, D.C., Krzeminski, W., Roth, M., & Rieke, M.J. 1998, AJ, 116, 2475

Johnson, H.L. 1966, ARAA, 4, 193

Tokunaga, A. 2000, Infrared Astronomy, in Allen's Astrophysical Quantities, 4th edition, A.N. Cox, Springer-Verlag: NY, 143

Tokunaga, A. T., Simons, D. A., & Vacca, W.D. 2001, submitted for publication

Wainscoat, R. J. & Cowie, L. L. 1992, AJ, 103, 332

van der Bliek, N.S., Manfroid, J., & Bouchet, P. 1996, A&AS, 119, 547

Young, A. T., Milone, E. F., & Stagg, C. R. 1994, A&AS, 105, 259

Figure Captions

Fig. 1 – A representative model bandpass plotted with Mauna Kea atmospheric transmission.

Fig. 2 – Transmission vs. airmass in the *K* band is shown. Crosses denote model transmission of the filter and atmosphere shown in Figure 1 for various air masses. The line is the least-squares fit.

Fig. 3 – Model sky emission above Mauna Kea is shown for the *K* window. A combination of OH lines and thermal flux defines the emission. In addition, telescope contributions, assuming either 3% (thick line) or 10% (thin line) emissivity are shown.

Fig. 4 – The atmospheric *K* window is plotted together with the Young et al. 1994 "improved *K*" filter, which has a triangular shape and has been tuned to deliver optimal photometric accuracy, albeit at the expense of throughput. Also plotted are scans of standard Barr *K* and *K´* filters, derived from CFHT Redeye filter data.

Fig. 5 – Results of *J* band modeling showing atmospheric transmission for the (top) standard Barr and (middle) MKO-NIR filter. The bottom panel shows an estimate of the photometric linearity $\sigma$ and $\tau_0$ as a function of filter bandpass for Mauna Kea and a 2 km altitude site.

Fig. 6 – Results of *H* band modeling are summarized.

Fig. 7 – Results of *K* band modeling are summarized.

Fig. 8 – Bandpasses of special purpose *K* band filters are shown.

Fig. 9 – Results of *L´* band modeling are summarized.

Fig. 10 – For comparison the bandpasses of Barr *L* and *M* filters is shown.

Fig. 11 – Results of *M´* band modeling are summarized.

Table 1. Adopted Central Wavelengths for the MKO Filter Set

| Bandpass | $\lambda_c$ (µm) |
|:---:|:---:|
| J | 1.24 |
| H | 1.65 |
| K | 2.20 |
| L´ | 3.77 |
| M´ | 4.67 |

Table 2. Bandpasses for Stock Infrared Filters

| Filter | Barr $\lambda_1 - \lambda_2$ (μm) | UKIRT $\lambda_1 - \lambda_2$ (μm) | IRTF $\lambda_1 - \lambda_2$ (μm) | CFHT $\lambda_1 - \lambda_2$ (μm) |
|---|---|---|---|---|
| J | 1.11-1.39 | 1.13-1.42 | 1.11-1.42 | 1.10-1.39 |
| H | 1.50-1.80 | 1.53-1.81 | 1.48-1.76 | 1.51-1.79 |
| K | 2.00-2.40 | 2.00-2.41 | 2.02-2.41 | 2.02-2.41 |
| K´ | | | 1.95-2.29 | 1.95-2.29 |
| $K_s$ | | 1.99-2.32 | 1.99-2.31 | |
| L | 3.20-3.80 | 3.15-3.75 | 3.20-3.81 | |
| L´ | 3.50-4.10 | 3.50-4.10 | 3.49-4.08 | |
| M | 4.50-5.10 | | 4.54-5.16 | |
| M´ | | 4.55-4.80 | 4.67-4.89 | |

Table 3. Comparative Summary of Filter Sets for Mauna Kea and Mid-Altitude (2 km high) Site

| | | MKO-NIR Filters | | | | | | Barr Filters | | | | |
| | | Mauna Kea | | | 2 km | | | Mauna Kea | | | 2 km | |
| Filter | Bandpass[a] (μm) | $\tau_0$ (%) | σ (millimag) | Sky (γ s$^{-1}$ arcsec$^{-2}$) | $\tau_0$ (%) | σ (millimag) | Bandpass[a] (μm) | $\tau_0$ (%) | σ (millimag) | Sky (γ s$^{-1}$ arcsec$^{-2}$) | $\tau_0$ (%) | σ (millimag) |
|---|---|---|---|---|---|---|---|---|---|---|---|---|
| J | 1.17 – 1.33 | 76 | 1.0 | ~4e4[b] | 77 | 1.6 | 1.11 – 1.39 | 66 | 4.6 | ~8e4[c] | 65 | 5.1 |
| H | 1.49 – 1.78 | 79 | 0.6 | 4.8e5 | 79 | 0.9 | 1.50 – 1.80 | 78 | 0.9 | 4.9e5 | 78 | 1.5 |
| K | 2.03 – 2.37 | 74 | 1.2 | 1.0e5 | 75 | 1.6 | 2.00 – 2.40 | 73 | 2.0 | 1.3e5 | 73 | 2.8 |
| K′ | ... | ... | ... | ... | ... | ... | 1.95 – 2.29[c] | 70[c] | 2.7[c] | 1.2e5[c] | 70[c] | 3.8[c] |
| $K_s$ | ... | ... | ... | ... | ... | ... | 1.99 – 2.31[c] | 72[c] | 1.9[c] | 1.0e5[c] | 72[c] | 2.8[c] |
| $K_l$ | 2.07 – 2.41 | 74 | 1.1 | 1.0e5 | 76 | 1.4 | ... | ... | ... | ... | ... | ... |
| L | ... | ... | ... | ... | ... | ... | 3.20 – 3.80 | 63 | 5.2 | 7.5e7 | 61 | 5.7 |
| L′ | 3.42 – 4.12 | 73 | 1.4 | 1.4e8 | 68 | 3.7 | 3.50 – 4.10 | 72 | 1.2 | 1.1e8 | 68 | 3.1 |
| M | ... | ... | ... | ... | ... | ... | 4.50 – 5.10 | 53 | 8.5 | 2.1e9 | 43 | 12.1 |
| M′ | 4.57 – 4.79 | 55 | 5.9 | 3.8e8 | 46 | 8.1 | 4.67 – 4.89[c] | 45[c] | 7.8[c] | 5.6e8[c] | 38[c] | 10.0[c] |

[a]Denotes ~50% transmission points in filter profile

[b]Estimate based upon Redeye measurements at CFHT

[c]IRTF bandpass used.

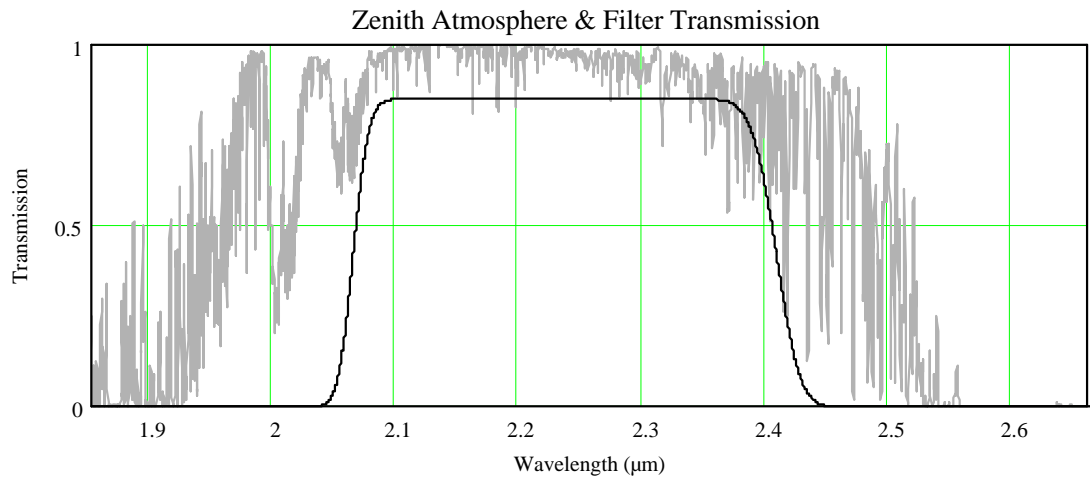

Figure 1

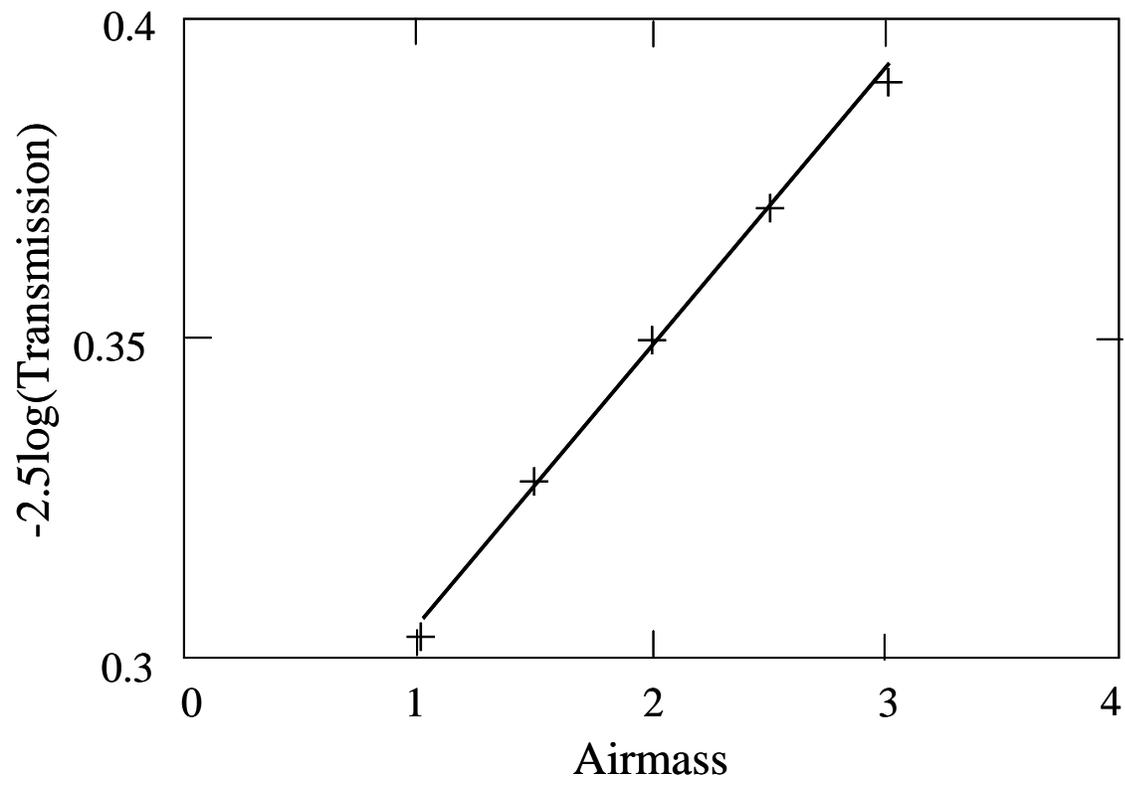

Figure 2

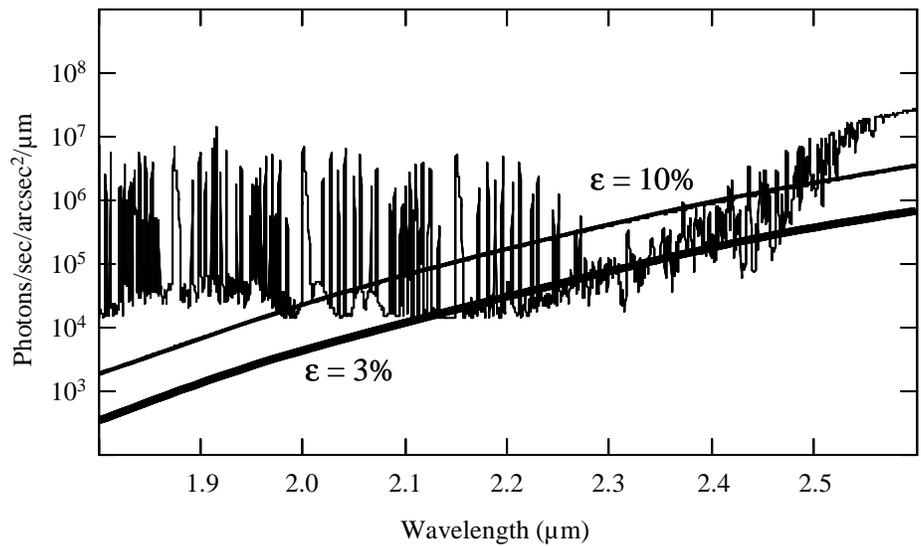

Figure 3

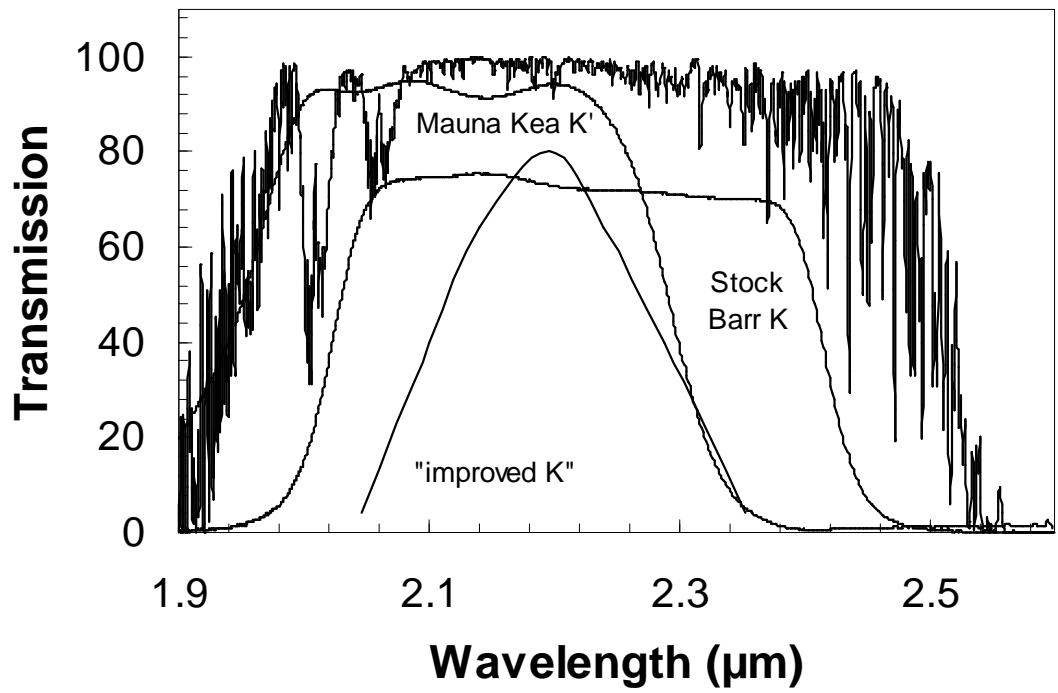

Figure 4

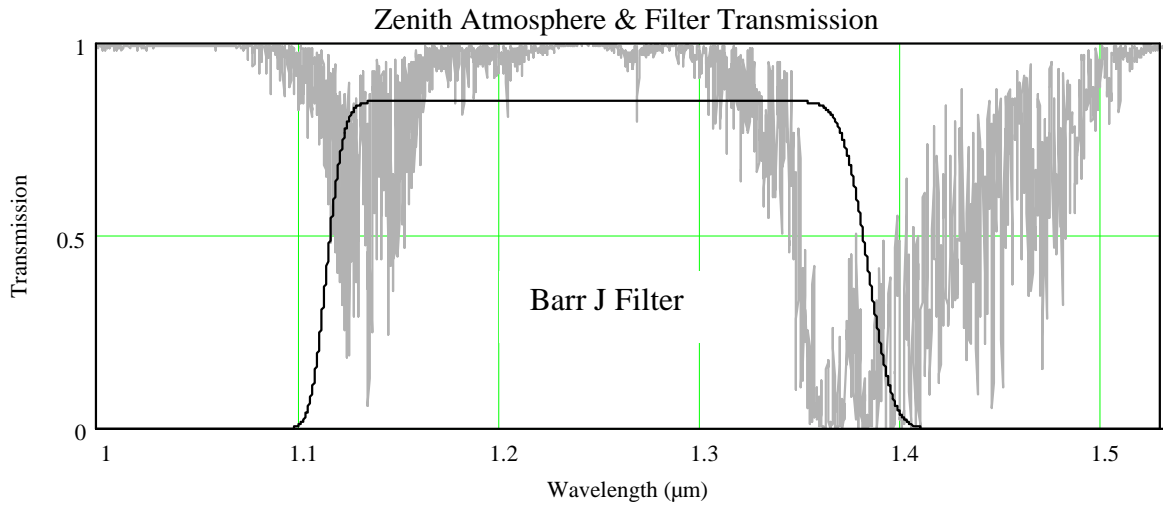

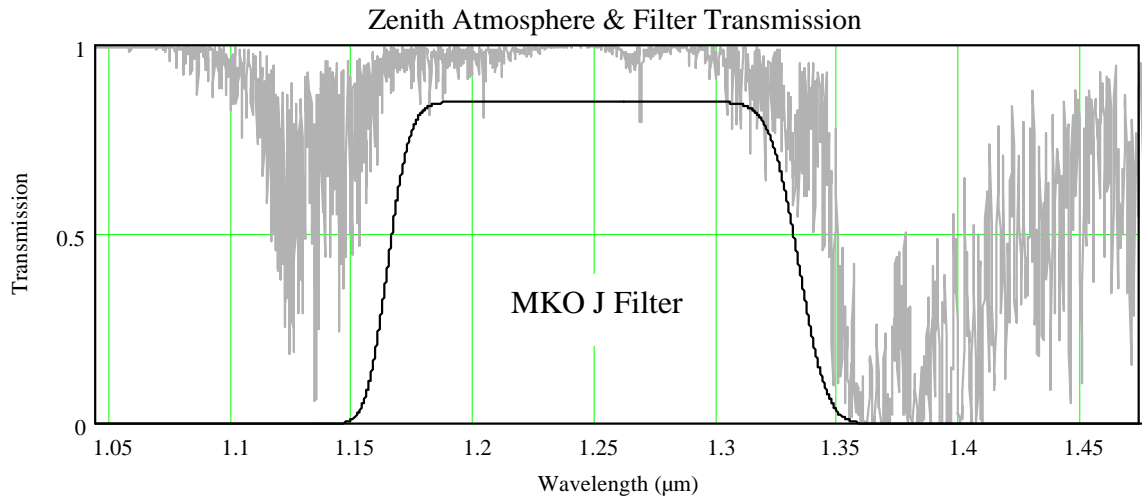

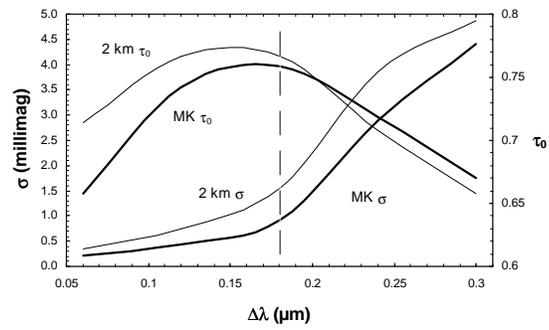

Figure 5

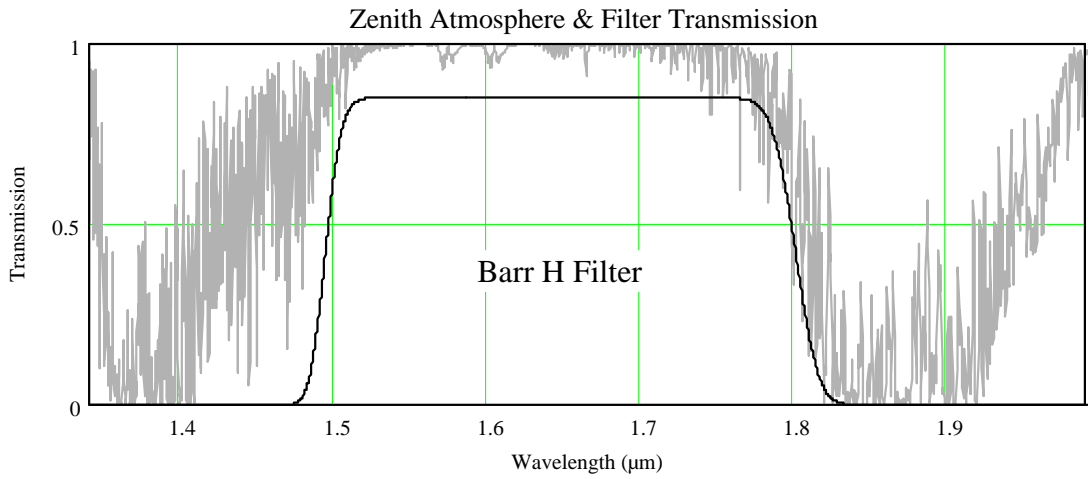
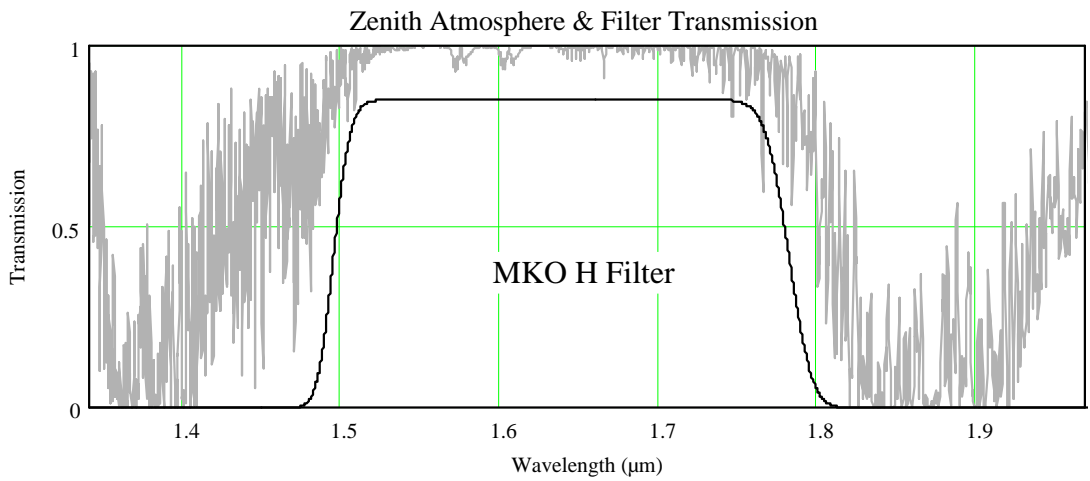
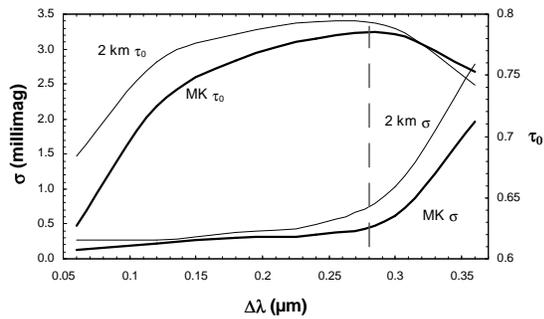
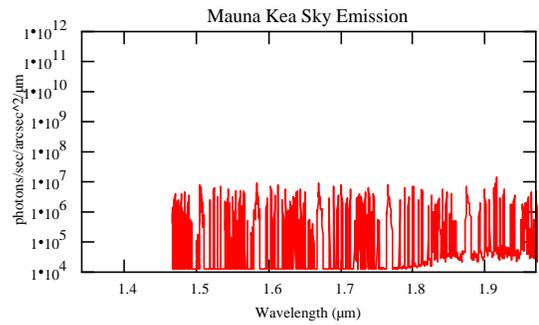

Figure 6

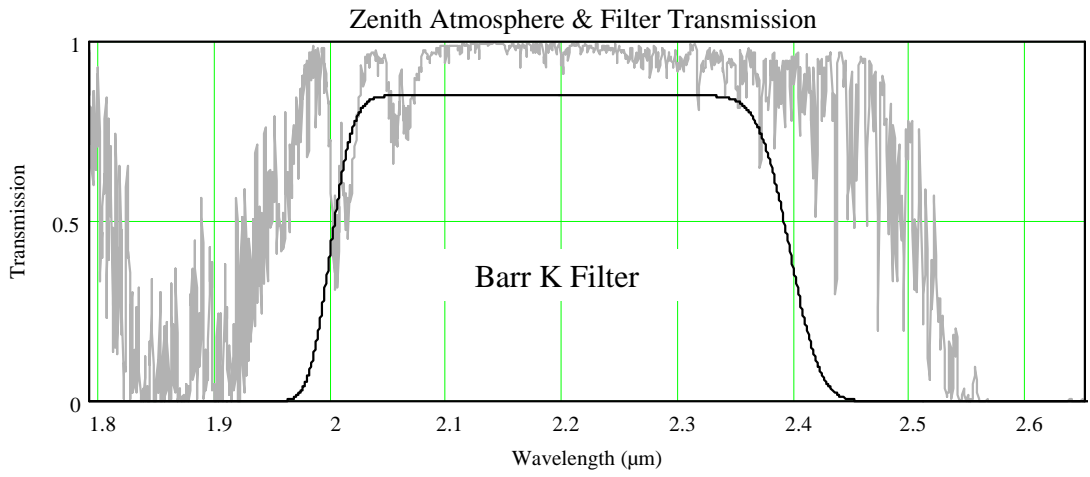

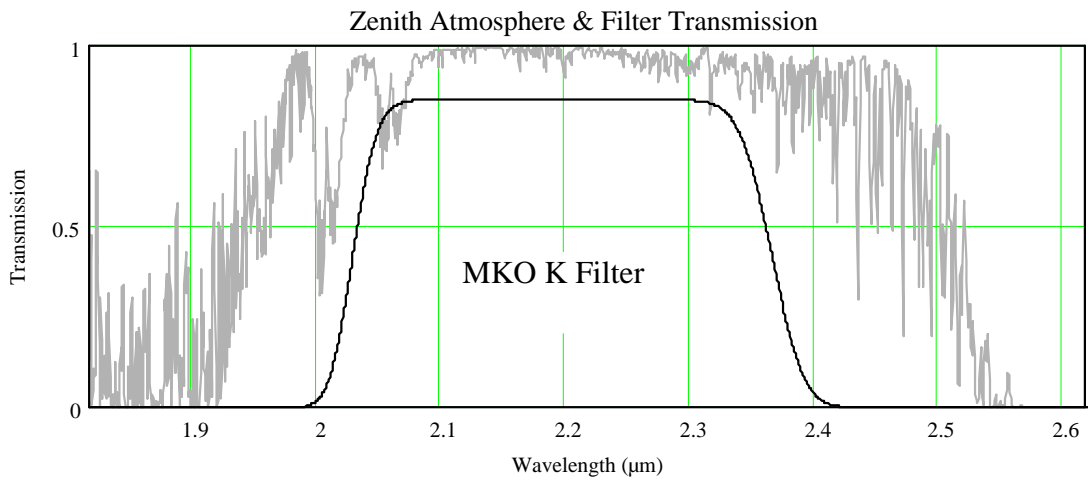

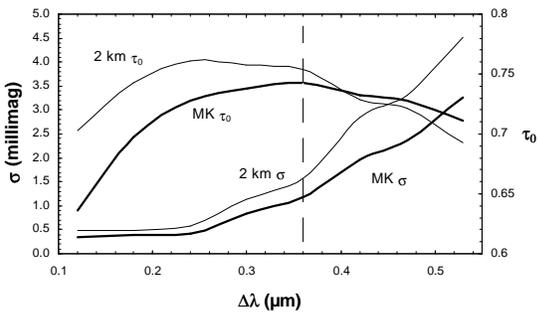
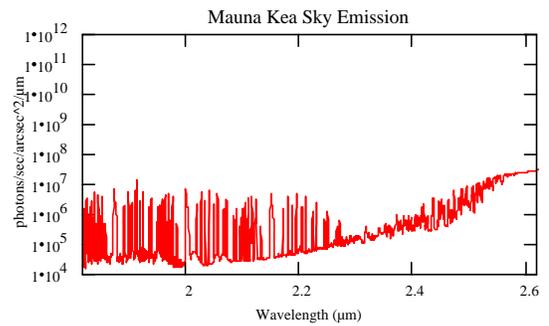

Figure 7

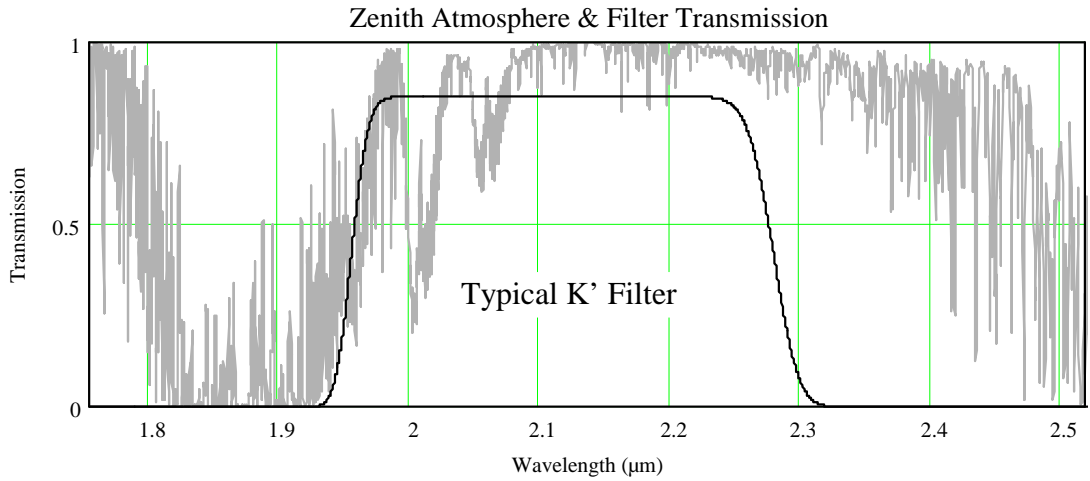

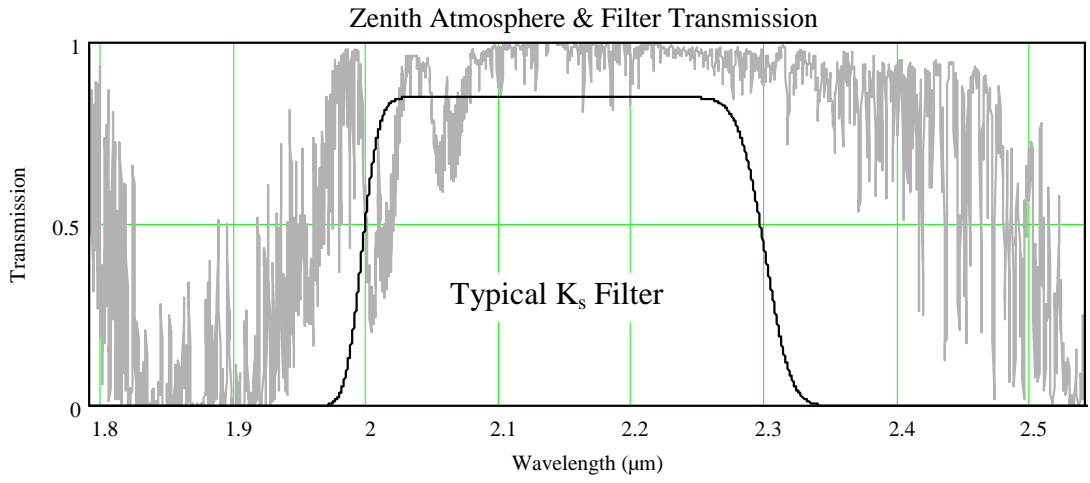

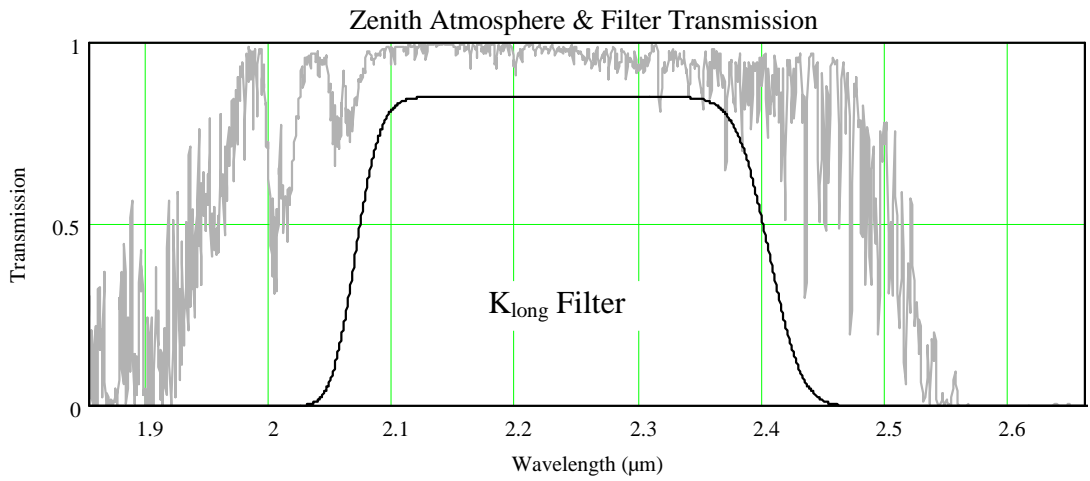

Figure 8

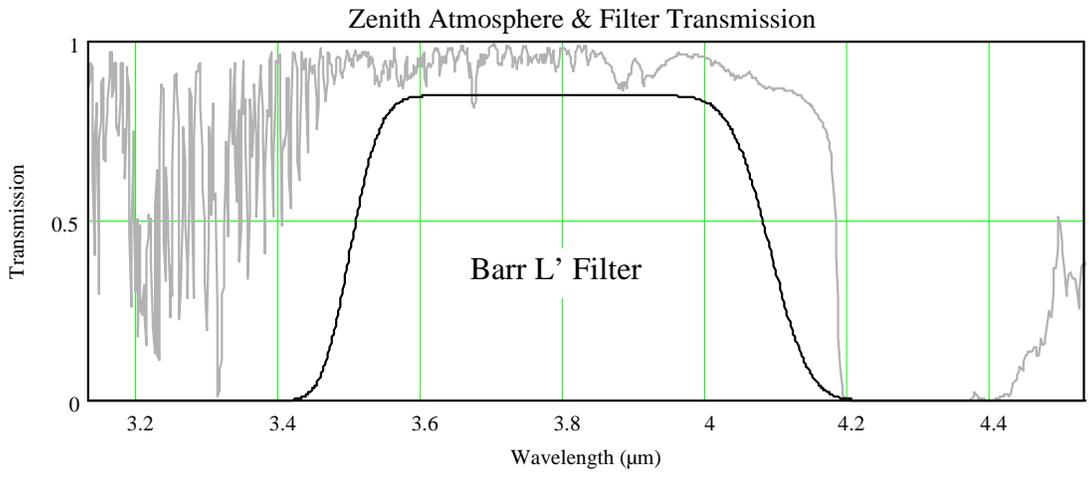

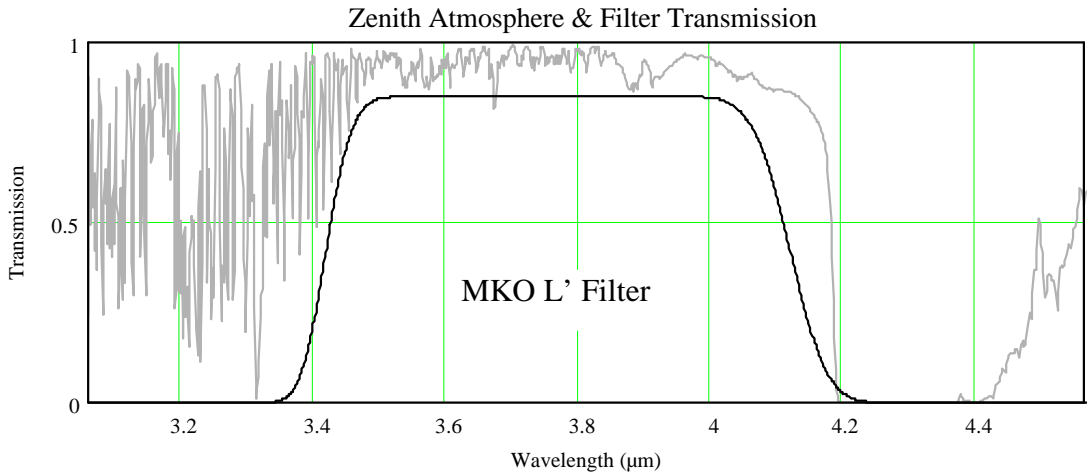

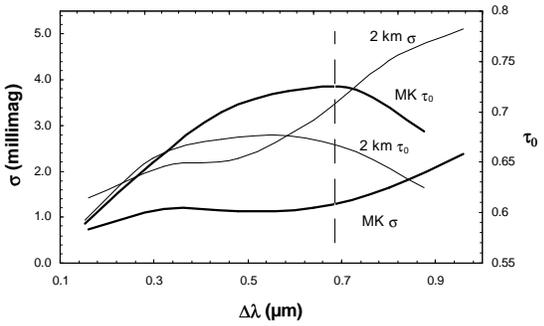

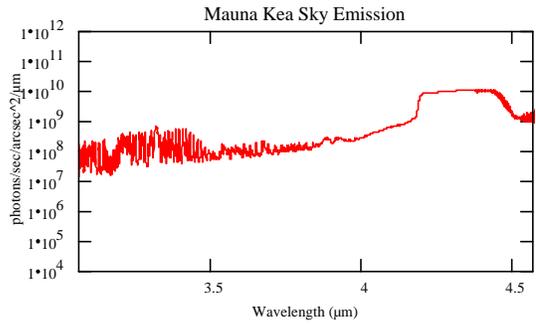

Figure 9

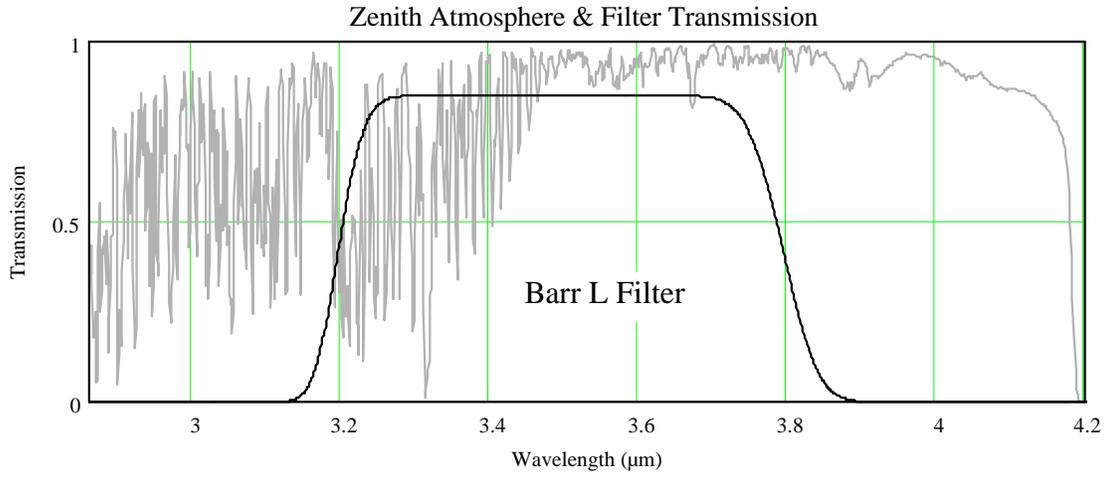

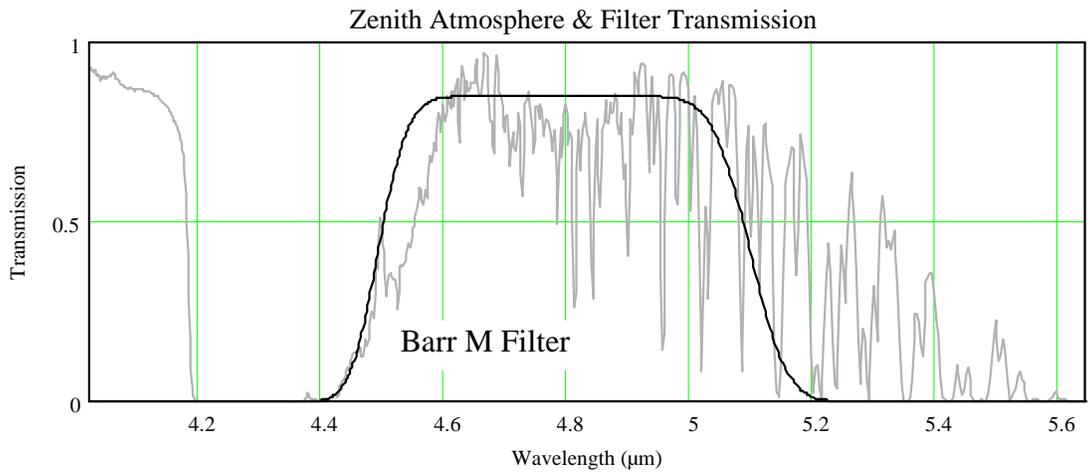

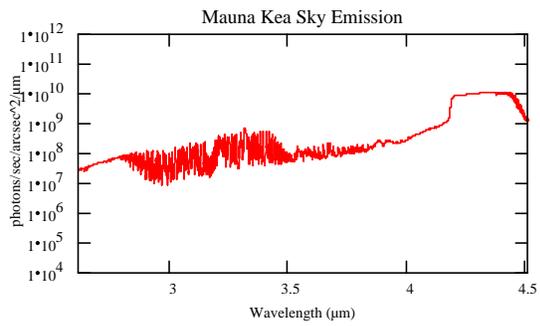
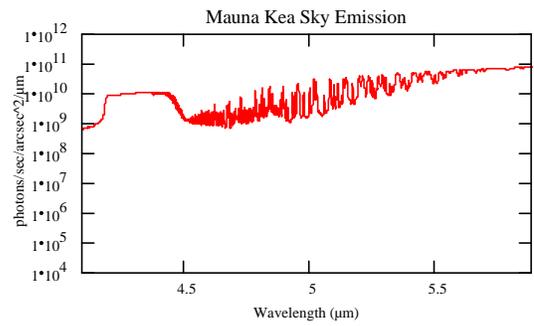

Figure 10

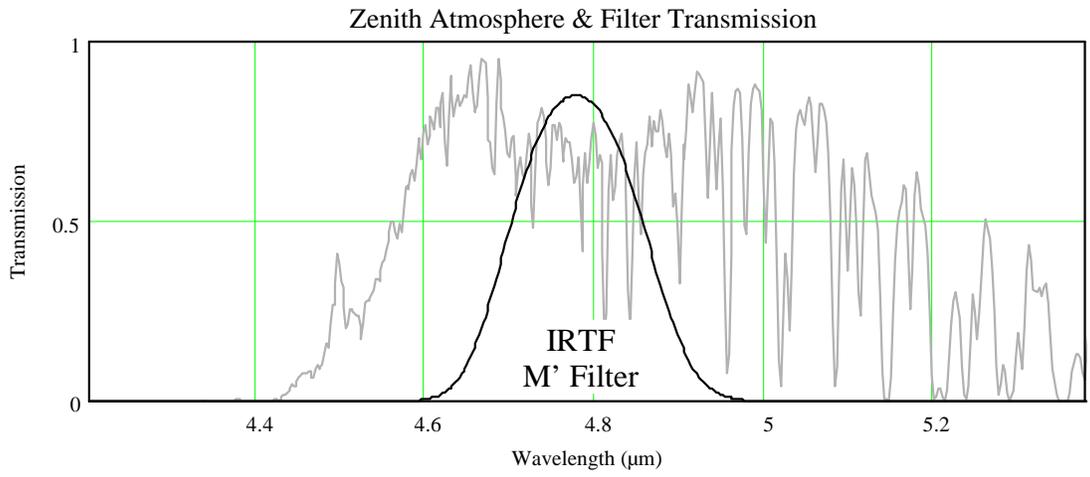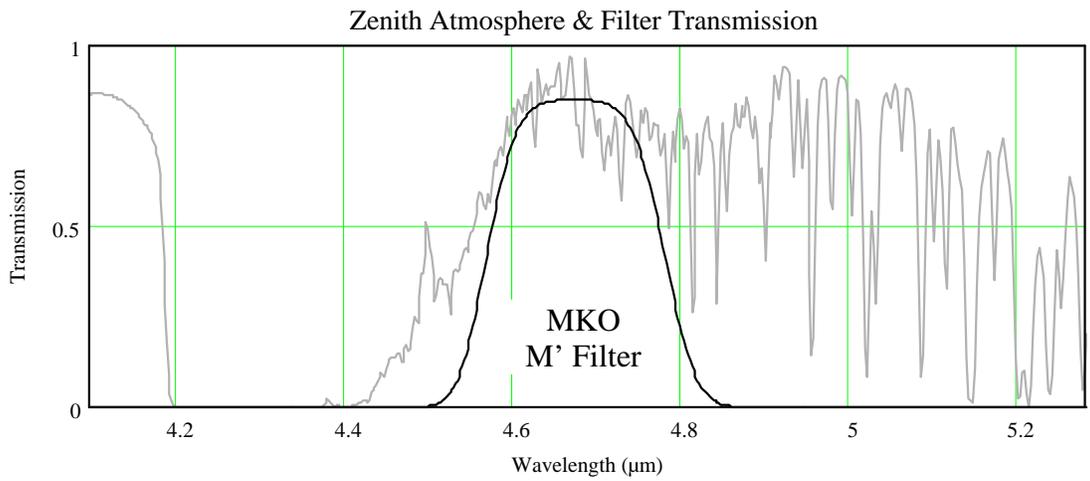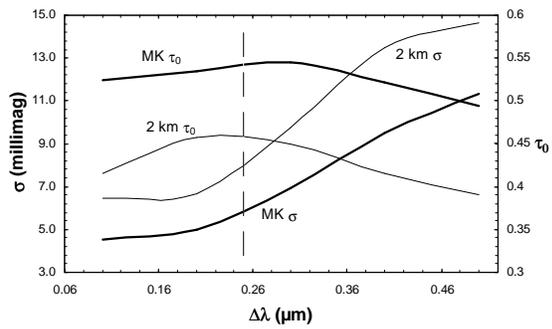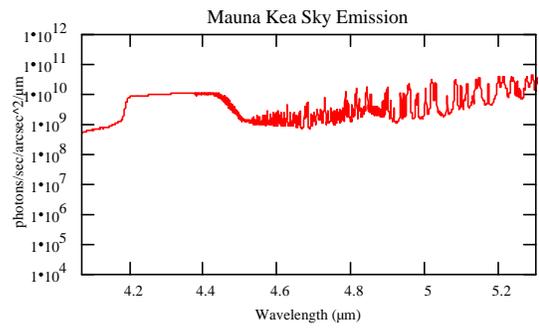

Figure 11